\documentclass[a4paper,dvips,12pt]{article}
\usepackage{amsmath,latexsym,amssymb,epsfig,psfrag}  
\usepackage{axodraw} 
\usepackage{comment}
\usepackage{cite}
\usepackage{wasysym}

\newcommand{\be}{\begin{equation}}
\newcommand{\ee}{\end{equation}}

\newcommand{\barra}{\,/\!\!\!}
\def\bc{\begin{center}}
\def\ec{\end{center}}
\def\bi{\begin{itemize}}
\def\ei{\end{itemize}}
\def\bea{\begin{eqnarray}}
\def\eea{\end{eqnarray}}

\def\nn{\nonumber}
\makeatletter   
\@addtoreset{equation}{section}   
\makeatother

\def\simlt{\stackrel{<}{{}_\sim}}
\def\simgt{\stackrel{>}{{}_\sim}}  
\begin{document}  
  
\baselineskip=16pt   
\begin{titlepage}  
\begin{center}  
\hfill{{\bf UAB-FT-571}}

\vspace{2cm}  
  
\large {\sc \Large Higgs-gauge unification without tadpoles}

\vspace*{10mm}  
\normalsize  
  
{\bf C.~Biggio~\footnote{biggio@ifae.es} and
M.~Quir\'os~\footnote{quiros@ifae.es}}   

\vspace*{5mm}  

\smallskip   
\medskip   
\it{~$^{1,2}$~Theoretical Physics Group, IFAE/UAB}\\ 
\it{E-08193 Bellaterra (Barcelona) Spain}

\smallskip   
\medskip   
\it{~$^{2}$~Instituci\'o Catalana de Recerca i Estudis Avan\c{c}ats (ICREA)}

\vskip0.6in \end{center}  
   
\centerline{\large\bf Abstract}  
\vspace*{10mm} 
\noindent
In orbifold gauge theories localized tadpoles can be radiatively
generated at the fixed points where $U(1)$ subgroups are conserved. If
the Standard Model Higgs fields are identified with internal
components of the bulk gauge fields (Higgs-gauge unification) in the
presence of these tadpoles the Higgs mass becomes sensitive to the UV
cutoff and electroweak symmetry breaking is spoiled. We find the
general conditions, based on symmetry arguments, for the
absence/presence of localized tadpoles in models with an arbitrary
number of dimensions $D$. We show that in the class of orbifold
compactifications based on $T^{D-4}/\mathbb Z_N$ ($D$ even, $N>2$)
tadpoles are always allowed, while on $T^{D-4}/\mathbb Z_2$ (arbitrary
$D$) with fermions in arbitrary representations of the bulk gauge
group tadpoles can only appear in $D=6$ dimensions. We explicitly
check this with one- and two-loops calculations.

\vspace*{2mm}

\end{titlepage}  
  
\section{\sc Introduction}
\label{Intro}
The Standard Model (SM) of strong, weak and electromagnetic
interactions is currently considered as an effective theory below a
given cutoff $\Lambda_{\rm SM}$. While this cutoff should be
$\Lambda_{\rm EW}\simlt 1$~TeV for the stability of the Higgs mass
under radiative corrections, present bounds from the non-observation
of (dimension-six) four-fermion operators~\cite{PDG} push the SM
cutoff towards $\Lambda\simgt$~10~TeV. This order of magnitude
discrepancy between the cutoff $\Lambda_{\rm EW}$ (required for the
stability of the electroweak symmetry breaking) and the cutoff
$\Lambda$ (implied by SM precision tests) already requires a certain
amount of fine-tuning that is known as the little hierarchy
problem~\cite{Giudice:2003nc}.

Supersymmetry remains of course the best solution to the little (and
grand) hierarchy problem, providing a SM cutoff of the order of the
mass of supersymmetric partners $\Lambda_{\rm EW}\sim M_{\rm
SUSY}$. The little hierarchy problem is naturally solved if $R$-parity
is conserved: in this case supersymmetric virtual effects are loop
suppressed and $\Lambda_{\rm EW}\sim 4\pi\Lambda$.  Nevertheless the
minimal supersymmetric SM extension (the MSSM) is becoming very
constrained by the LEP bounds from Higgs searches for radiative
corrections to the Higgs quartic coupling increase only
logarithmically with the scale $M_{\rm SUSY}$ which is a source of
fine-tuning in the MSSM. It is thus compelling to propose possible
alternative solutions to the little hierarchy problem that could fill
the gap between the sub-TeV scale required for the stability of the
electroweak symmetry breaking and the multi-TeV scale required by
precision tests of the SM.

The possibility of TeV extra dimensions~\cite{Antoniadis:1990ew},
suggested by string theories~\cite{Lykken:1993br}, has motivated an
alternative solution to the little hierarchy problem called
Higgs-gauge
unification~\cite{Randjbar-Daemi:1982hi,Antoniadis:1993jp,Hatanaka:1999sx,
Antoniadis:2000tq,ABQ,vonGersdorff:2002us,
Csaki:2002ur,Scrucca:2003ut} in which the internal components of
higher dimensional gauge bosons play the role of the SM Higgses that
acquire a non-vanishing vacuum expectation value (VEV) through the
Hosotani mechanism~\cite{Hosotani:1983xw,Hosotani:2004ka}. In this
framework the Higgs mass in the bulk is protected from quadratic
divergences by the higher-dimensional gauge theory and only finite
corrections $\propto (1/R)^2$ ($R$ is the compactification radius)
that disappear in the $R\to\infty$ limit can appear. The SM cutoff is
then identified with the inverse compactification radius $1/R$. On the
other hand since the higher dimensional theory is non-renormalizable,
it is in turn an effective theory with a cutoff $\Lambda$. The little
hierarchy between $1/R$ and $\Lambda$ is protected by the
higher-dimensional gauge theory.  Of course localized terms can be
generated at the orbifold fixed points by quantum
corrections~\cite{ggh,Contino:2001si,vonGersdorff:2002as} consistently
with the symmetries of the
theory~\cite{vonGersdorff:2002rg,vonGersdorff:2002us}. A direct
localized squared mass ($\sim\Lambda^2$) for the Higgs-gauge fields is
forbidden by a shift symmetry at the orbifold fixed points: a remnant
of the original bulk gauge symmetry~\cite{vonGersdorff:2002rg}.

It was however realized~\cite{vonGersdorff:2002us,Csaki:2002ur} that
this approach could be jeopardized by the radiative generation of a
localized tadpole in the cases where the bulk gauge group is broken at
the orbifold fixed points into a group containing $U(1)$ factors, as
e.g.~the hypercharge $U(1)$. The tadpole corresponds to the field
strength for internal space dimensions along the $U(1)$-direction and
contains in its non-abelian part a term quadratic in the Higgs-gauge
fields. In this way the tadpole quadratic divergence amounts to a
quadratic divergence for the Higgs-gauge field squared mass and spoils
the little hierarchy we wanted to solve. The existence of such
tadpoles is a generic feature of orbifold compactifications in
dimensions $D\geq 6$ and has been confirmed in six-dimensional
orbifold field~\cite{Scrucca:2003ut} and ten-dimensional
string~\cite{GrootNibbelink:2003gb} theories. One way
out~\cite{Scrucca:2003ut} is that local tadpoles vanish globally and
thus they would not spoil the four dimensional effective theory,
although this requires a strong restriction on the bulk fermion
content. Another possibility, that will be explored in this work, is
that localized tadpoles be inconsistent with the symmetries of the
theory in which case they will not be radiatively generated.

In this paper we will find the general conditions for the absence of
localized tadpoles at the orbifold fixed points where $U(1)$ subgroups
are conserved.  They depend on the particular subgroup of the internal
tangent space group $SO(D-4)$ which remains unbroken at the fixed
point, and which constitutes a global invariance of the localized
Lagrangian. In particular, if the internal rotation group acting on
the component $A_i$ is $SO(p)$ ($p>2$), the absence of tadpoles
involving that particular component is guaranteed, since only p-forms
$F_{i_1\dots i_p}$ can be linearly generated as $\epsilon^{i_1\dots
i_p}F_{i_1\dots i_p}$, where $\epsilon^{i_1\dots i_p}$ is the
corresponding Levi-Civita tensor. Conversely if $p=2$ the $U(1)$
tadpole can be generated through the gauge invariant form
$\epsilon^{ij}F_{ij}$.  The knowledge of such general conditions
should enable us to find orbifold field theories with electroweak
symmetry breaking triggered by Higgs-gauge fields where the Higgs mass
is insensitive to the UV cutoff, although we will not attempt here to
construct realistic tadpole-free models. Instead we will present a
general class of such models based on $\mathbb Z_2$ orbifolds. In fact
we will find that in the general class of $T^{D-4}/ \mathbb Z_2$
orbifold compactifications tadpoles only appear in the case of $D=6$
extra dimensions. For any other dimension the theory is
tadpole-free. For $D\neq 6$ tadpoles are forbidden by the symmetries
of the theory and we have explicitly checked this cancellation at the
one- and two-loop levels. On the contrary we show that in the class of
orbifold compactifications $T^{D-4}/ \mathbb Z_N$ ($D$ even, $N>2$)
tadpoles are always allowed.

The rest of this paper is organized as follows. In
Section~\ref{general} we present the general arguments based on
symmetry considerations for the absence of localized tadpoles. In
particular it is shown in Section~\ref{Z2} that the general class of
compactifications based on $T^{D-4}/\mathbb Z_2$ for $D>6$ fulfills
the general requirements for the absence of localized tadpoles.  In
subsections~\ref{one} and~\ref{two} we show by explicit one- and
two-loop calculations respectively that indeed no tadpoles appear in
$T^{D-4}/\mathbb Z_2$ for $D\neq 6$. Section~\ref{Conclusions}
contains some conclusive remarks while technical details about traces
of $\Gamma$-matrices in arbitrary dimensions are presented in
Appendix~\ref{Gamma}.

\section{\sc General symmetry arguments}
\label{general}

Given a compact $d$-dimensional $(d=D-4)$ manifold $\mathcal C$ and a
discrete symmetry group $\mathbb G$ acting non-freely on it (i.e.~with
fixed points) we can define an orbifold by modding out $\mathcal C$ by
$\mathbb G$~\cite{orbifold}. We will consider the case $\mathcal
C=T^d$ where $T^d$ is the $d$-dimensional torus obtained by modding
out $\mathbb R^d$ by a $d$-dimensional lattice $\Lambda^d$: the
orbifold group is then generated by a discrete subgroup of $SO(d)$
that acts crystallographically on the torus lattice and by discrete
shifts that belong to the torus lattice. In particular if we define
the coordinates as $x^M=(x^\mu,y^i)$, where $x^\mu$ are
four-dimensional coordinates and $y^i$ $(i=1,\dots,d)$ the torus
coordinates, the action of $k\in\mathbb G$ on the torus is $k\cdot
y=P_k\, y+u$ and the inverse is defined as $k^{-1}\cdot y=P_k^{-1}\,
y-P_k^{-1}\,u$, where $P_k$ is a discrete rotation in $SO(d)$ and
$u\in \Lambda^d$; $y$ and $k\cdot y$ are then identified on the
orbifold. Since the orbifold group is acting non-freely on the torus
there are fixed points characterized by $k\cdot y_f=y_f$. Any given
fixed point $y_f$ remains invariant under the action of a subgroup
$\mathbb G_f$ of the orbifold group.

The orbifold group acts on fields $\phi_{\mathcal R}$ transforming as
an irreducible representation ${\mathcal R}$ of the gauge group
$\mathcal G$ as
\begin{equation}
k\cdot\phi_{\mathcal R}(y)=\lambda^k_{\mathcal R}\otimes\mathcal P^k_\sigma\,
\phi_{\mathcal R}(k^{-1}\cdot y)
\label{orbfield}
\end{equation}
where $\lambda^k_{\mathcal R}$ is acting on gauge and flavor indices
and $\mathcal P^k_\sigma$, where $\sigma$ refers to the field spin, on
Lorentz indices. In particular one finds for scalar fields $\mathcal
P_0^k=1$ and for gauge fields $\mathcal P_1^k=P_k$ for a discrete
rotation ($\mathcal P_1^k=1$ for a lattice shift) that trivially
follows from the invariance of the one-form $A=A_i dy^i$. Similarly
the field strength $F=F_{ij}dy^i\wedge dy^j$ transforms as $\mathcal
P^k_2 F=P_k F P_k^T$.  For fermions $\mathcal{P}^k_{\frac{1}{2}}$ can
be derived requiring the invariance of the lagrangian under the
orbifold action, as we will discuss in more detail later on.  On the
other side $\lambda^k_{\mathcal R}$ depends on the gauge structure and
the gauge breaking of the orbifold action. $\lambda^k_{\mathcal R}$
and $\mathcal P^k_\sigma$ are representations of the orbifold element
$k$ on the gauge and Lorentz groups, respectively.

In general the orbifold action breaks the gauge group in the bulk
$\mathcal G=\{T^A\}$ to a subgroup $\mathcal H_f=\{T^{a_f}\}$ at the
fixed point $y_f$. The orbifold group element $k$ acts as a Lie
algebra automorphism $T^A\to \Lambda_k^{AB}T^B$ that can be
represented as a group conjugation $T^A\to g_k T^A g_k^{-1}$ in case
of an inner automorphism (as we will consider here), where $g_k$ is a
representation on $\mathcal G$ of the orbifold group element $k$. A
convenient basis to write the element $g_k$ is the Weyl-Cartan basis
of the Lie algebra $\{T^A\}=\{E^\alpha,H^I\}$, where $\vec\alpha$ is
the rank$(\mathcal G)$-dimensional root associated to the generator
$E^\alpha$. In this case the gauge group breaking is characterized by
the vector $\vec v^{\,k}$ as
\begin{equation}
\lambda^k_{\mathcal R}=\exp(-2\pi i \,\vec v^{\,k}\cdot \vec H_{\mathcal R})
\end{equation}
and indeed one can identify $g^k=\lambda^k_{Adj}$ and the Lie algebra
automorphism is given by $\Lambda_k={\rm diag}[\delta_{IJ},\exp(-2\pi
i \vec{v}^{\,k}\cdot\vec\alpha)\delta_{\alpha\beta}]$. The group
elements $\lambda^k_{\mathcal R}$ satisfy the automorphism condition
$\lambda^k_{\mathcal R} T^A_{\mathcal R}\lambda^{-k}_{\mathcal
R}=\Lambda^{AB}T^B_{\mathcal R}$, from where it follows that the
subgroup $\mathcal H_f$ left invariant by the orbifold elements
$k\in\mathbb G_f$ is defined by the generators that commute with
$\lambda^k_{\mathcal R}$, i.e.~$[\lambda^k_{\mathcal
R},T^{a_f}_{\mathcal R}]=0$. Of course the latter condition must be
satisfied by any irreducible representation $\mathcal R$ of $\mathcal
G$.

In the same way as the orbifold action breaks the bulk gauge group
$\mathcal G$ to a subgroup $\mathcal H_f$, such that
$[\lambda^k,\mathcal H_f]=0$, it also breaks the internal rotation
group $SO(d)$ into a subgroup $\mathcal O_f$ at the orbifold fixed
point $y_f$. In fact in compactifications to a smooth $d$-dimensional
Riemannian manifold (with positive signature) the orthogonal
transformations acting on the tangent space at a given point form the
group $SO(d)$~\cite{Witten}. At the orbifold fixed point $y_f$ a
further compatibility condition between the orbifold action and the
internal rotations is required. In particular, if the given fixed
point $y_f$ is left invariant by the orbifold subgroup $\mathbb G_f$,
only $\mathbb G_f$-invariant operators $\Phi_{\mathcal R,\sigma}$
(either invariant fields, products of non-invariant fields or
derivatives of fields) couple to $y_f$,~i.e.
\begin{equation}
k\cdot\Phi_{\mathcal R,\sigma}(y_f)=\Phi_{\mathcal R,\sigma}(y_f)\ .
\end{equation}
Acting on $\Phi_{\mathcal R,\sigma}$ with an internal rotation we get
a transformed operator that should also be $\mathbb
G_f$-invariant. This means, using Eq.~(\ref{orbfield}), that the
subgroup $\mathcal O_f$ is spanned by the generators of $SO(d)$ that
commute with $\mathcal P^k_{\sigma}$, i.e.~they satisfy the condition
\begin{equation}
[\mathcal O_f, \mathcal P^k_{\sigma}]=0
\label{conmutador}
\end{equation}
for $k\in\mathbb G_f$ and arbitrary values of $\sigma$.  In particular
in the presence of gauge fields $A_M=(A_\mu,A_i)$ an invariant
operator can be $F_{ij}$ with $\mathcal R=Adj$ and $\sigma=2$.  The
internal components $A_i$ transform under the action of the orbifold
element $k\in \mathbb G_f$ as the discrete rotation $P_k$. At the
orbifold fixed point $y_f$ only the subgroup $\mathcal O_f\subseteq
SO(d)$ survives and the vector representation $A_i$ of $SO(d)$ breaks
into irreducible representations of $\mathcal O_f$.

We now consider a gauge theory coupled to fermions. The lagrangian of
the orbifold theory is the sum of a bulk $D$-dimensional lagrangian
$\mathcal L_D$ and four-dimensional lagrangians $\mathcal L_f$
localized at the orbifold fixed points $y=y_f$ as
\begin{equation}
\mathcal L=\mathcal L_D+\sum_{f}\delta^{(d)}(y-y_f)\,\mathcal L_f \, .
\end{equation}
The bulk lagrangian is given by
\be
\label{lagr}
\mathcal{L}_D= -\frac{1}{4} F^A_{MN} F^{A MN} + i\bar{\Psi}\Gamma_D^M D_M
\Psi 
\ee 
with $F^A_{MN}=\partial_M A_N^A - \partial_N A_M^A - g f^{ABC} A_M^B
A_N^C$, $D_M=\partial_M-igA_M^AT^A$ and where $\Gamma_D^M$ are the
$\Gamma$-matrices corresponding to a $D$-dimensional space-time that
are defined in appendix~\ref{Gamma}. The four-dimensional lagrangians
$\mathcal L_f$ must be invariant under the usual four-dimensional
symmetries: gauge [$\mathcal H_f$] and Lorentz [$SO(3,1)$]
symmetries. On top of that they must be invariant under the action of
the orbifold group and the remnant symmetries left out by the orbifold
compactification: the remnant gauge symmetry and the internal rotation
group $\mathcal O_f$. We will now briefly comment about the two latter
symmetries. Notice that they are global symmetries of the Lagrangian
$\mathcal L_f$.

The invariance under bulk (infinitesimal) gauge transformations
\begin{equation}
\delta_\xi A_M^A=\frac{1}{g}\partial_M\xi^A-f^{ABC}\xi^BA_M^C,\quad
\delta_\xi \Psi=i\xi^AT^A\Psi
\label{gtrans}
\end{equation}
translates, when applied to the orbifold fixed points $y_f$, into the
four-dimensional gauge symmetry~\footnote{From here on we will remove
for simplicity the subscript ``f'' from the gauge group and the
corresponding generators.}  $\mathcal H=\{T^a\}$ that applies to the
four-dimensional gauge fields $A_\mu^a$ which are also invariant under
the orbifold action and leads to the usual gauge invariance under
$\mathcal H$-transformations $\delta_\xi
A_\mu^a=\partial_\mu\xi^a/g-f^{abc}\xi^bA_\mu^c$. By localizing the
transformations (\ref{gtrans}) at the orbifold fixed point $y_f$ and
keeping the orbifold invariant terms one can define an infinite set of
transformations (remnant of the bulk gauge invariance) induced by
derivatives of $\xi^A$ that we can call $\mathcal
K$-transformations~\cite{vonGersdorff:2002rg,vonGersdorff:2002us}. Then
only $\mathcal H$ and $\mathcal K$-invariant quantities are allowed at
the orbifold fixed points. For instance if the gauge field $A^{\hat
a}_{i}$ is invariant under the orbifold action, where $T^{\hat
a}\in\mathcal G/\mathcal H$, the remnant ``shift'' symmetry
$\delta_\xi A_{i}^{\hat a}=\partial_{i}\xi^{\hat a}/g+\dots$ prevents
the corresponding zero mode from acquiring a mass localized at the
orbifold fixed point.

Gauge fields along the internal dimensions $A_i^a$ are scalars in the
adjoint representation of the group $\mathcal H$ that transform under
the orbifold action as $P_k A$; they are not orbifold invariant and
cannot couple to fixed points. However the corresponding field
strength $F_{ij}^a$ transforms under the orbifold action as $k\cdot
F^a_{ij}=(P_k)_{i}^m(P_k)_{j}^nF^a_{mn}$ and, depending on the
orbifold, some components $F^{a}_{ij}$ can be orbifold invariant. On
the other hand gauge fields along internal dimensions and components
in the coset $\mathcal G/\mathcal H=\{T^{\hat a}\}$ transform under
the orbifold group as $k\cdot A_i^{\hat a}=(P_k)_{i}^j\Lambda_k^{\hat
a B}A_j^B$: some components $A_i^{\hat a}$ can be orbifold invariant.
Under these circumstances if $\mathcal H\supseteq U(1)$ the
``tadpole'' $F_{ij}^\alpha$ where $\alpha$ is the $U(1)$ quantum
number
\begin{equation}
F_{ij}^\alpha=\partial_i A_j^\alpha-\partial_jA_i^\alpha-gf^{\alpha \hat
b \hat c}A_i^{\hat b}A_j^{\hat c}
\label{fij}
\end{equation}
is $\mathcal H$ and $\mathcal K$-invariant as well as orbifold
invariant. Notice that the (orbifold invariant part of the) last term
can appear as a mass term for zero modes.

In summary if $F_{ij}^\alpha$ appears localized at orbifold fixed
points it can contain a zero mode mass term that can destabilize
(break) the gauge theory~\footnote{We assume here that $U(1)$ is not
contained in the bulk group $\mathcal G$.  Otherwise the non-abelian
term in (\ref{fij}) does not exist and the tadpole is harmless as far
as the electroweak breaking is concerned.}.  For orbifold
compactifications breaking $\mathcal G$ into $U(1)$ subgroups at the
fixed points the existence of gauge and orbifold-invariant field
strengths $F_{ij}^\alpha$ that can appear in localized lagrangians is
a generic feature in any model. The further requirement for the
tadpole appearance is that internal rotation invariance be conserved
at the fixed point.

We have previously seen that the vector representation $A_i$ of
$SO(d)$ breaks into irreducible representations of the internal
rotation group $\mathcal O_f\subseteq SO(d)$ that commutes with
$\mathcal P^k_\sigma$. In particular if the rotation subgroup acting
on the ($i,j$)-indices is $SO(2)$ then $\epsilon^{ij}F^\alpha_{ij}$,
where $\epsilon^{ij}$ is the Levi-Civita tensor, is invariant under
$\mathcal O_f$. On the other hand if the rotation subgroup acting on
the ($i,j$)-indices is $SO(p)$ ($p>2$) then the Levi-Civita tensor
would be $\epsilon^{i_1 i_2\dots i_p}$ and only invariants constructed
using p-forms would be allowed. In other words a sufficient condition
for the absence of localized tadpoles is that the smallest internal
subgroup factor be $SO(p)$ ($p>2$).

We can exemplify the different possibilities by considering a class of
$\mathbb G=\mathbb Z_N$ orbifolds for even $d$ where the generator
$P_N$ of the orbifold group is defined by
\begin{equation}
P_N=\prod_{i=1}^{d/2} e^{2\pi i\frac{k_i}{N}J_{2i-1,2i}}
\label{pe}
\end{equation}
where $k_i$ are integer numbers ($0<k_i<N$) and $J_{2i-1,2i}$ is the
generator of a rotation with angle $2\pi\frac{k_i}{N}$ in the plane
$(y_{2i-1},y_{2i})$. All orbifold elements are defined by $P_k=P_N^k$
$(k=1,\dots,N-1)$ and satisfy the condition $P_N^N=1$.  The generator
of rotations in the $(y_{2i-1},y_{2i})$-plane can be written as
$J_{2i-1,2i}={\rm diag}(0,\dots,\sigma^2,\dots,0)$ where the Pauli
matrix $\sigma^2$ is in the $i$-th two-by-two block. The generator
$P_N$ can be written as $P_N={\rm diag}(R_1,\dots,R_{d/2})$ where the
discrete rotation in the $(y_{2i-1},y_{2i})$-plane is defined as
\begin{equation}
R_i=\left(
\begin{array}{cc}
c_i & s_i\\
-s_i & c_i
\end{array}
\right)
\end{equation}
with $c_i=\cos(2\pi k_i/N)$, $s_i=\sin(2\pi k_i/N)$.

Let $y_f$ be a fixed point that is left invariant under the orbifold
subgroup $\mathbb G_f=\mathbb Z_{N_f}$ where $N_f\leq N$.  We now
define the internal rotation group $\mathcal O_f$ as the subgroup of
$SO(d)$ that commutes with the generator of the orbifold $\mathbb
Z_{N_f}$, $P_{N_f}$ as given by Eq.~(\ref{pe}) with $N$ replaced by $N_f$. In
general, if $N_f>2$ it is trivially provided by the tensor product:
\begin{equation}
\mathcal O_f=\bigotimes_{i=1}^{d/2}SO(2)_i
\label{grupo}
\end{equation}
where $SO(2)_i$ is the $SO(2)\subseteq SO(d)$ that acts on the
$(y_{2i-1},y_{2i})$-subspace. In every such subspace the metric is
$\delta_{IJ}$ and the Levi-Civita (antisymmetric) tensor
$\epsilon^{IJ}$ ($I,J=2i-1,2i$, $i=1,\dots,d/2$) such that we expect
the tadpoles appearance at the fixed points $y_f$ as
\begin{equation}
\sum_{i=1}^{d/2}\mathcal
C_i\sum_{I,J=2i-1}^{2i}\epsilon^{IJ}F^\alpha_{IJ}\;\delta^{(d/2)}(y-y_f)\, .
\label{tadpolos}
\end{equation}
If $N_f=2$ then the generator of the orbifold subgroup $\mathbb
G_f=\mathbb Z_2$ is the inversion $P=-\mathbf{1}$ that obviously commutes
with all generators of $SO(d)$ and $\mathcal O_f=SO(d)$. In this case
the Levi-Civita tensor is $\epsilon^{i_1\dots i_d}$ and only a d-form
can be generated linearly in the localized lagrangian. Therefore
tadpoles are only expected in the case of $d=2$ $(D=6)$.

The last comments also apply to the case of $\mathbb Z_2$ orbifolds of
arbitrary dimensions (even or odd) since in that case the orbifold
generator is always $P=-\mathbf{1}$ and the internal rotation group
that commutes with $P$ is $\mathcal O_f=SO(d)$ for all the fixed
points. Again tadpoles are only expected for $D=6$ dimensions while
they should not appear for $D>6$~\footnote{Notice that tadpoles vanish
identically for the case $D=5$.}.

Since every operator in $\mathcal L_f$ that is consistent with all the
symmetries should be radiatively generated by loop effects from matter
in the bulk (unless it is protected by some other $-$accidental$-$
symmetry) explicit calculations of the tadpole in orbifold gauge
theories should confirm the appearance (or absence) of them in
agreement with the above symmetry arguments. In the rest of this paper
we will explicitly present the case of $\mathbb{Z}_2$-orbifolds in
arbitrary dimensions.

\section{\sc $\mathbb{Z}_2$ orbifolds in arbitrary dimension}
\label{Z2}

We consider in this section the case of a gauge theory coupled to
fermions in $D>4$ dimensions. The bulk lagrangian is given in
Eq.~(\ref{lagr}) and the extra dimensions are compactified on the
orbifold $T^d/\mathbb Z_2$, with the action of $\mathbb Z_2$ defined
by the identification $y^i=-y^i$. The orbifold group $\mathbb Z_2$ has
a single element (apart from the identity) i.e.~$P=-\mathbf{1}$.

The parity assignment for gauge fields is characterized by the
diagonal matrix $\Lambda={\rm diag}(\eta^A)$ with $\eta^A=\pm 1$. It can
then be written as:
\be
\label{gaugeZ2}
A^A_M(x^\mu,-y^i) = \eta^A \alpha_M A^A_M(x^\mu,y^i) \ ,
\ee 
with $\alpha_\mu=+1$, $\alpha_i=-1$, $\eta^a=+1$ and
$\eta^{\hat{a}}=-1$. Here $a$ corresponds to the unbroken generators
of the gauge group, while $\hat{a}$ corresponds to the broken ones.
The only conditions we need for the Yang-Mills term to be invariant
under this $\mathbb Z_2$ action is the automorphism condition
\cite{Hebecker:2001jb,vonGersdorff:2002as}
\be
\label{gaugecond}
\eta^A\eta^B\eta^C=1 \textrm{  for  } f^{ABC}\ne 0 \ .
\ee

The action of $\mathbb Z_2$ on fermions in representation ${\mathcal
R}$ of the gauge group $\mathcal G$ is
\be
\label{fermionZ2}
\Psi_{\mathcal R}(x^\mu,-y^i) = 
\lambda_{\mathcal R} \otimes \mathcal P_{\frac{1}{2}} 
\Psi_{\mathcal R}(x^\mu,y^i)\ ,
\ee 
where $\lambda_{\mathcal R}$ acts on representation indices and
$\mathcal P_{\frac{1}{2}}$ on spinor indices. From the requirement
that the kinetic term for $\Psi_{\mathcal R}$ is invariant under the
parity action we obtain the following constraint on $\mathcal
P_{\frac{1}{2}}$:
\be
\label{cond1}
\Gamma_D^0 \mathcal P_{\frac{1}{2}} \Gamma_D^0 \Gamma_D^M = 
\alpha_M \Gamma_D^M  \mathcal P_{\frac{1}{2}} \ ,
\ee
which translates into two possible different conditions:
\be
\label{cond2}
\begin{array}{llll}
\textrm{if }\left[\Gamma_D^0,\mathcal P_{\frac{1}{2}}\right]=0 & \Rightarrow &
\left\{
\begin{array}{l}
\left[\Gamma_D^{\mu},\mathcal P_{\frac{1}{2}}\right]=0\\[0.2cm]
\left\{\Gamma_D^i,\mathcal P_{\frac{1}{2}}\right\}=0
\end{array}
\right.  &\ \ (a) \\[0.5cm] \textrm{if }\left\{\Gamma_D^0,\mathcal
P_{\frac{1}{2}}\right\}=0 & \Rightarrow & \left\{
\begin{array}{l}
\left\{\Gamma_D^{\mu},\mathcal P_{\frac{1}{2}}\right\}=0\\[0.2cm]
\left[\Gamma_D^i,\mathcal P_{\frac{1}{2}}\right]=0 \ .
\end{array}
\right. &\ \ (b)
\end{array}
\ee
All this is valid for every $D$. It can be shown that for $D$ even
there is a solution in both cases, while for $D$ odd only ($b$) has a
solution. These are precisely:
\be
\label{P1/2}
\begin{array}{ll}
D \textrm{ even} &\left\{
\begin{array}{ll}
(a)\Rightarrow & \mathcal P_{\frac{1}{2}}=\beta_D\ \Gamma_D^5...\Gamma_D^D
\\[0.2cm] (b)\Rightarrow & \mathcal P^\prime_{\frac{1}{2}}=\beta_D\
\Gamma_D^5...\Gamma_D^D \Gamma_D^{D+1}= \mathcal P_{\frac{1}{2}} \Gamma_D^{D+1}
\end{array}\right.\\[0.5cm]
D \textrm{ odd} &\left\{
\begin{array}{ll}
(a)\Rightarrow & \textrm{no solution}\\[0.2cm] (b)\Rightarrow &
\mathcal P^\prime_{\frac{1}{2}}=-i\beta_{D-1}\ \Gamma_D^5...\Gamma_D^D
\, ,
\end{array}\right.
\end{array}
\ee
where $\beta_D$ is such that $\mathcal P_{\frac{1}{2}}^2=\mathbf{1}$
(and therefore $\mathcal P_{\frac{1}{2}}^{\prime\,2}=\mathbf{1}$). Up
to now we have considered only the kinetic term; from the requirement
that also the interaction term is invariant under the parity
transformation, we obtain the following conditions on
$\lambda_{\mathcal R}$ \cite{vonGersdorff:2002as}:
\be
\label{cond3}
\eta^A \lambda_{\mathcal R} T^A  \lambda_{\mathcal R} 
= T^A \qquad\Leftrightarrow\qquad \left\{
\begin{array}{l}
\left[\lambda_{\mathcal R},T^a\right]=0\\[0.2cm]
\left\{\lambda_{\mathcal R},T^{\hat{a}}\right\}=0 \ .
\end{array}\right.
\ee
These requirements are valid for any $D$.

\subsection{\sc One-loop calculation of tadpoles}
\label{one}

Since we want to compute radiative corrections, we must define the
Feynman rules. It is well known that in an orbifold field theory all
the information concerning the orbifold can be inserted in the
propagator of the KK-modes of the fields, leaving the vertices
momentum-conserving~\cite{ggh}. In this picture the propagator of the
$\vec{m}$-mode ($\vec{m}=(m_1,...,m_d)$) of an arbitrary field $\Phi$
(a gauge boson $A_M$, a ghost field $c$ or a fermion field $\Psi$) in
the $D$ dimensional space compactified on $T^d/\mathbb Z_2$ can be
written in terms of the propagator of a field living in the torus
$T^d$ in the following way:
\be
\label{propagator}
\left< \Phi^{\vec{m}'} \bar{\Phi}^{\vec{m}} \right> =
\frac{1}{2} G^{\Phi}(p_\mu,p_i) 
(\delta_{\vec{m}'-\vec{m}} \pm \mathcal P_{\Phi} \delta_{\vec{m}'+\vec{m}}) \ .
\ee
Here $G^{\Phi}(p_\mu,p_i)$ is the propagator of $\Phi$ in a $D$
dimensional Minkowski space where the torus periodicity conditions
$p_i=m_i/R$ are imposed~\footnote{We are assuming a common
compactification radius $R$ for all internal dimensions and orthogonal
lattice vectors.}, $\mathcal P_{\Phi}$ is the parity action defined in
Eqs.~(\ref{gaugeZ2})-(\ref{fermionZ2}) and the ``$-$'' sign only
applies to fermions if $\mathcal P_{\frac{1}{2}}$ anticommutes with
$\Gamma_D^0$. The propagators in a $D$ dimensional Minkowski
space-time and in the Feynman gauge are the following:
\bea
\label{props}
G^{A}(p_\mu,p_i)&=& -i\,\frac{\delta^{BC}}{p^2}
\ g_{MN} \nn \\
G^{c}(p_\mu,p_i)&=& -i\,\frac{\delta^{BC}}{p^2}\nn \\
G^{\Psi}(p_\mu,p_i)&=& i\,\frac{\Gamma_{D}^M p_M}{p^2}\, ,
\eea
where $p^2\equiv p^M p_M$ is the $D$-dimensional momentum.
The vertices are given by (\ref{lagr}) and by the gauge fixing
and ghost lagrangian. In the Feynman gauge the latter is given by
\begin{equation}
\mathcal L_{\rm gf}+\mathcal L_{FP}=-\frac{1}{2}\left(
g^{MN}\partial_M A_N^A\right)^2-tr\,\bar c\;\partial^M D_M c \, .
\label{gf}
\end{equation}

\subsubsection{\sc Fermion contribution} 
\begin{figure}[htb]
\begin{center}
\SetScale{1.}
\begin{picture}(50,100)(25,0)
\ArrowArc(50,75)(25,-90,270)
\LongArrowArcn(50,75)(33,195,160)
\Text(10,75)[]{$q$}
\Photon(50,0)(50,50)36
\LongArrow(60,20)(60,40)
\Text(70,30)[]{$p$}
\Text(33,30)[]{$A_M^A$}
\end{picture}
\end{center}
\caption{One-loop fermion contribution to the tadpole}
\label{tadpolef}
\end{figure}
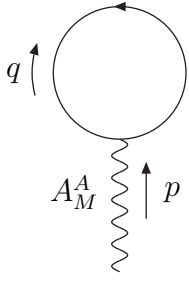

We begin by considering $D$ even in which case fermions can be
chiral. The one-loop fermion contribution to the tadpole $\partial_i
A_j^a$ is given by the diagram of Fig.~\ref{tadpolef} where the
fermionic line contains a projector $\mathcal P_{\Psi}$ coming from
the expansion in Eq.~(\ref{propagator}).

The fermion contribution to the tadpole turns out to be proportional to:
\be
\label{tadfermeven1}
tr\left\{\frac{\Gamma_D^N q_N}{q^2}
\left(\delta_{\vec{m}'-\vec{m}} \pm \lambda{_\mathcal R} \otimes \mathcal
P_{\frac{1}{2}} \delta_{\vec{m}^\prime+\vec{m}}\right)
\delta_{\vec{m}^\prime-\vec{m}+\vec{l}} \left(\mathbf{1} \pm
\Gamma_D^{D+1} \right) \Gamma_D^M T^A_{\mathcal R} \right\} \ .  
\ee
The term proportional to $\delta_{\vec{m}'-\vec{m}}$ which corresponds
to the bulk contribution is identically zero since it contains
$tr\{T^A_{\mathcal R}\}$ which vanishes.  We consider then the brane
contribution, which can be factorized as follows:
\be
\label{tadfermeven2}
tr\left\{\Gamma_D^N \mathcal P_{\frac{1}{2}}
\left(\mathbf{1} \pm \Gamma_D^{D+1} \right) \Gamma_D^M \right\}
tr\left\{\lambda_{\mathcal R} T^A_{\mathcal R} \right\}
\frac{q_N}{q^2} \ .
\ee
By simply considering the trace relative to the $\Gamma$-matrices, it
is possible to show that Eq.~(\ref{tadfermeven2}) can be non-zero only
for $D=6$. Indeed this term can be written as
\be
\label{tadfermeven3}
tr\left\{\Gamma_D^N \mathcal P_{\frac{1}{2}} \left(\mathbf{1} \pm
\Gamma_D^{D+1} \right) \Gamma_D^M \right\}= tr\left\{\mathcal
P_{\frac{1}{2}} \Gamma_D^M \Gamma_D^N \right\} \pm tr\left\{\mathcal
P_{\frac{1}{2}} \Gamma_D^{D+1} \Gamma_D^M \Gamma_D^N \right\}\ , 
\ee
where $\mathcal P_{\frac{1}{2}}$ is one of the two solutions of
Eq.~(\ref{P1/2}). If $\mathcal P_{\frac{1}{2}}\propto
\Gamma_D^5\dots\Gamma_D^D$ (case $a$), using the rules on the trace
enumerated in the appendix, it can be shown that only the first term
of the right-hand side of Eq.~(\ref{tadfermeven3}) can be different
from zero and this only happens for $D=6$. In this particular case the
result is:
\be
\label{trace1a}
tr\left\{\Gamma_D^5 \Gamma_D^6 \Gamma_D^M \Gamma_D^N \right\}=
8 (g^{5N}g^{6M}-g^{5M}g^{6N}) = - 8 g^{Mi}g^{Nj}\epsilon_{ij}\, .
\ee
In case ($b$), where $\mathcal P^\prime_{\frac{1}{2}}= \mathcal
P_{\frac{1}{2}} \Gamma_D^{D+1}$, the traces we have to evaluate are
the same but inverted. This means that only the second term of the
right-hand side of Eq.~(\ref{tadfermeven3}) can be different from
zero, the result being the one quoted above. The only noticeable thing
is that while in case ($a$) the contribution to the tadpole was
chiral-independent, in case ($b$) it is chiral-dependent. This means
that in six dimensions if fermions transform under parity according to
$\mathcal P_{\frac{1}{2}}$ we have a non-vanishing tadpole both with
Dirac and Weyl fermions. On the contrary, if they transform with
$\mathcal P^\prime_{\frac{1}{2}}$, the tadpole can be zero if we are
in presence of Dirac fermions, since fermions with different chirality
give opposite contributions. This is a consequence of an extra parity
symmetry of the theory which inverts separately the internal
coordinates: the term $\epsilon^{ij}F_{ij}$ is odd under this symmetry
and therefore it cannot be
generated~\cite{vonGersdorff:2002us,Csaki:2002ur,Scrucca:2003ut}.

As already discussed in Ref.~\cite{vonGersdorff:2002us}, there is a
close relation between the tadpole generated by the diagram of
Fig.~\ref{tadpolef} and the mixed $U(1)$-gravitational anomaly in a
$6D$ theory. In fact fermions with different $6D$ chiralities
contribute with the same sign to the tadpole and with opposite sign to
the anomaly or viceversa, depending on the particular $\mathcal
P_{\frac{1}{2}}$ considered. As a consequence, starting with Dirac
fermions, a vanishing anomaly implies a non-vanishing tadpole and
viceversa. The cancellation conditions are the same for the tadpole
and the anomaly only when dealing with chiral fermions.

We now move to the case of $D$ odd. Here chirality does not exist so
the fermion contribution is proportional to:
\be
\label{tadfermodd1}
tr\left\{\frac{\Gamma_D^N q_N}{q^2}
\left(\delta_{\vec{m}'-\vec{m}} \pm \lambda_{\mathcal R} \otimes
\mathcal P_{\frac{1}{2}} \delta_{\vec{m}'+\vec{m}}\right)
\delta_{\vec{m}'-\vec{m}+\vec{l}}\; \Gamma_D^M T^A_{\mathcal R}
\right\} \ .  
\ee
This means that the only trace we have to evaluate is
\be
\label{tadfermodd2}
tr\left\{\mathcal P_{\frac{1}{2}} \Gamma_D^M \Gamma_D^N \right\}\ .
\ee
Now $\mathcal P_{\frac{1}{2}}$ is unique and proportional to
$\Gamma_D^5\dots\Gamma_D^D$, where the
$\Gamma_D^D=\Gamma_{D-1}^{(D-1)+1}$. It is not difficult to show that
this contribution is always zero for any $D$ odd.

\subsubsection{\sc Gauge and ghost contribution} 

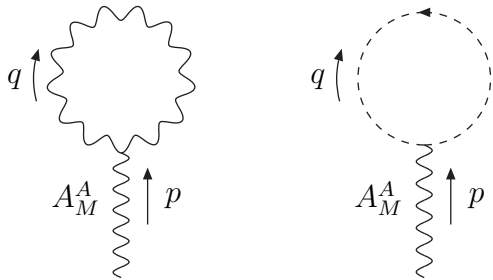
\begin{figure}[htb]
\begin{center}
\SetScale{1}
\begin{picture}(50,100)(25,0)
\PhotonArc(50,75)(25,14,374){3}{13}
\LongArrowArcn(50,75)(33,195,160)
\Text(10,75)[]{$q$}
\Photon(50,47)(50,0)36
\LongArrow(60,20)(60,40)
\Text(70,30)[]{$p$}
\Text(33,30)[]{$A_M^A$}
\end{picture}
\hspace{2cm}
\begin{picture}(50,100)(25,0)
\DashArrowArc(50,75)(25,-90,270)3
\LongArrowArcn(50,75)(33,195,160)
\Text(10,75)[]{$q$}
\Photon(50,0)(50,50)36
\LongArrow(60,20)(60,40)
\Text(70,30)[]{$p$}
\Text(33,30)[]{$A_M^A$}
\end{picture}
\end{center}
\caption{One-loop gauge and ghost contributions to the tadpole}
\label{tadpoleg}
\end{figure}

The gauge and ghost one-loop contributions to the tadpole $\partial_i A_j$ 
are given by the diagrams in Fig.~\ref{tadpoleg}.
They do not contribute to the tadpole for any dimension $D$. For $D=6$
the proof was done in Ref.~\cite{vonGersdorff:2002us} although the
cancellation is more general and could be applied to any
dimension. Indeed the gauge contribution is proportional to:
\be
\label{tadgauge2}
\delta^{BC} f^{ABC}
\ee
and this is clearly zero. The ghost contribution is also proportional
to this trace, so we can conclude that we do not have any contribution
from the gauge sector for any number of dimensions $D$. This is a
consequence of the same parity symmetry we previously discussed for
fermions. While there its existence depends on the particular
$\mathcal P_{\frac{1}{2}}$ considered, in the gauge sector it always
subsists, forbidding the appearance of the
tadpole~\cite{Csaki:2002ur,Scrucca:2003ut}.

\subsection{\sc Two-loop calculation}
\label{two}

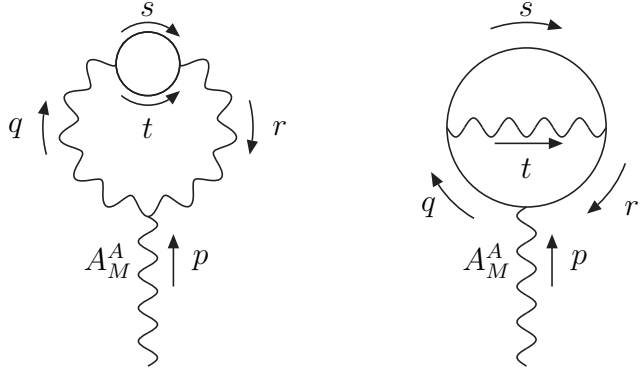
\begin{figure}[htb]
\begin{center}
\SetScale{1.2}
\hspace{1cm}
\begin{picture}(50,100)(0,0)
\PhotonArc(25,75)(25,14,374){3}{13}
\LongArrowArcn(25,75)(33,15,345)
\Text(80,90)[]{$r$}
\LongArrowArcn(25,75)(33,195,165)
\Text(30,90)[]{$t$}
\Text(30,135)[]{$s$}
\LongArrowArcn(25,95)(13,130,400)
\Text(-20,90)[]{$q$}
\LongArrowArc(25,95)(13,230,320)
\LongArrow(33,25)(33,40)
\Text(50,40)[]{$p$}
\Text(15,40)[]{$A_M^A$}
\BCirc(25,95){10}
\SetColor{Black}
\CArc(25,95)(10,-90,270)
\Photon(25,47)(25,0)3{4.5}
\end{picture}
\hspace{3cm}
\begin{picture}(50,100)(0,0)
\CArc(25,75)(25,-90,270)
\Photon(50,75)(0,75)3{4.5}
\Photon(25,0)(25,50)3{4.5}
\LongArrow(33,25)(33,40)
\Text(50,40)[]{$p$}
\Text(15,40)[]{$A_M^A$}
\LongArrowArcn(25,75)(33,-20,305)
\Text(70,60)[]{$r$}
\LongArrowArcn(25,75)(33,-120,205)
\Text(-7,60)[]{$q$}
\LongArrowArcn(25,75)(33,110,70)
\Text(30,135)[]{$s$}
\LongArrow(15,70)(35,70)
\Text(30,75)[]{$t$}
\end{picture}
\hspace{1cm}
\end{center}
\caption{Two-loop tadpole diagrams with fermions}
\label{twoloops}
\end{figure}

The cancellation of two-loop diagrams involving only gauge and ghost
fields was done in Ref.~\cite{vonGersdorff:2002us} for the case of
$D=6$ dimensions. Since the proof given there also applies for any
dimension we will skip it here. Two-loop diagrams involving fermion
loops (where fermions belong to the representation $\mathcal R$ of
$\mathcal G$) are shown in Fig.~\ref{twoloops} where all momenta are
$D$-dimensional and the external four-momentum is zero,
i.e.~$p=(0,p^i)$.\vspace{1.2cm}

The contribution of localized tadpoles corresponds to the given
diagrams with orbifold projections $\mathcal P_\Phi$ acting on some
(or all) internal lines.  In particular there is no localized
contribution only in the obvious case without any orbifold projections
and in the case in which there are two orbifold projections on the
legs connected to the external one. Anyway we will show that also in
these cases the (non-localized) result is zero as it should.  An
orbifold projection on the gauge boson propagator in Eq.~(\ref{props})
amounts to the insertion of $\eta^B\alpha_M$ and one on the fermion
propagator amounts to inserting $\lambda_{\mathcal R}\otimes\mathcal
P_{\frac{1}{2}}$. Since the orbifold projection is twisting the arrow
of internal momenta we will call an internal line with an orbifold
projection a ``twisted'' line.

Let us first discuss the diagram on the left of
Fig.~\ref{twoloops}. It is proportional to:
\begin{align}
&\sum_{\vec{q},\, \vec{s},\, \vec{q}^{\,\prime},\, \vec{s}^{\,\prime}}
\int\frac{d^4q}{(2\pi)^4}\frac{d^4s}{(2\pi)^4}
\frac{1}{q^2}\frac{1}{s^2}\frac{1}{t^2}\frac{1}{r^2}\,
tr\Big\{ \left(\delta_{\vec{q}-\vec{q}^{\,\prime}}
+\eta^B\alpha_N \delta_{\vec{q}+\vec{q}^{\,\prime}}\right)
\nonumber\\
&\Gamma_D^N \,T^B_{\mathcal R} 
\,\delta_{\vec{q}^{\,\prime}-\vec{s}-\vec{t}} \,/\!\!\!s 
\left(\delta_{\vec{s}-\vec{s}^{\,\prime}}+
\lambda_{\mathcal R}\otimes\mathcal P_{\frac{1}{2}}
\delta_{\vec{s}+\vec{s}^{\,\prime}}\right)\Gamma_D^R \,T^C_{\mathcal R}  
\,\delta_{\vec{s}^{\,\prime}+\vec{t}^{\,\prime}-\vec{r}^{\,\prime}} \,/\!\!\!t
\nonumber\\
&\left(\delta_{\vec{t}-\vec{t}^{\,\prime}}
+\lambda_{\mathcal R}\otimes\mathcal P_{\frac{1}{2}}
\delta_{\vec{t}+\vec{t}^{\,\prime}}\right) 
\left(\delta_{\vec{r}-\vec{r}^{\,\prime}}
+\eta^C\alpha_R \delta_{\vec{r}+\vec{r}^{\,\prime}}\right)
f^{ABC}
\nonumber\\
&\left[(r+q)_M g_{NR}-(q+p)_R g_{MN}+(p-r)_N g_{MR} \right]
\delta_{\vec{p}-\vec{q}+\vec{r}}\Big\}\, ,
\label{firstdiagram}
\end{align}
where $t^\mu=q^\mu-s^\mu$ and $r^\mu=q^\mu-p^\mu=q^\mu$ and we skip
the insertion of possible chirality factors
($\mathbf{1}\pm\Gamma_D^{D+1}$) for the case of chiral fermions in
even dimensions.  We see that there are sixteen different diagrams
which correspond to the different possibilities for inserting
projections, some of them differing only by permutations of twisted
lines. These sixteen diagrams can be divided into three groups
according to the number of twisted fermions: there are four diagrams
with no twisted fermions, eight with one twisted fermion and four with
two.

Diagrams with no twisted fermions are proportional to
\begin{equation}
f^{ABC} tr\left( T^B_{\mathcal R} T^C_{\mathcal R}\right)=
f^{ABC} \delta^{BC}C_2(\mathcal R)=0 
\label{first-zerotf}
\end{equation}
and so they cancel for any dimension. The presence or not of twisted
bosons does not affect this result, since it only consists in
multiplying this by $\eta^{B,C}$.

Diagrams with two twisted fermions are proportional to
\begin{equation}
f^{ABC} tr\left( T^B_{\mathcal R} \lambda_{\mathcal R}
T^C_{\mathcal R}\lambda_{\mathcal R}\right)= 
f^{ABC}\eta^C\delta^{BC}C_2(\mathcal R)=0
\label{first-twotf}
\end{equation}
and so they equally cancel for any dimension. 

Therefore we only have to compute diagrams with one twisted
fermion. Considering all the possible permutations of twists we have
eight diagrams, with zero, one or two twisted gauge bosons. The
diagrams in which the momentum $s$ is twisted are proportional to
\be
tr\left\{\Gamma_D^N\,/\!\!\!s\,\mathcal
P_{\frac{1}{2}}\Gamma_D^R\,/\!\!\!t\,\right\}\,
\left[(r+q)_M g_{NR}-(q+p)_R g_{MN}+(p-r)_N g_{MR} \right]
\label{first-s0}
\ee
and the presence of twisted gauge bosons only amounts to the insertion
of $\eta^{B,C}$ and, obviously, to a change in the flux of momenta.
Now we evaluate the above expression in the case of $M=i$, which
corresponds to have a gauge boson $A_i$ on the external leg. Indeed
the only quadratically divergent term that can be generated at the
orbifold fixed points consistently with the gauge symmetry is
$F_{ij}^\alpha$, so we can concentrate ourselves in isolating
contributions to $\partial_j A_i$ that precisely correspond to $M=i$
and an external momentum $p^j$. We obtain:
\begin{align}
&tr\left\{\Gamma_D^N \Gamma_D^S \mathcal P_{\frac{1}{2}}
\Gamma_D^R \Gamma_D^T\right\} s_S\,  t_T\,  (r+q)_i\,  g_{NR}\, -
\nonumber\\
&tr\left\{\Gamma_D^i \Gamma_D^S \mathcal P_{\frac{1}{2}}
\Gamma_D^R \Gamma_D^T\right\} s_S\,  t_T\,  (q+p)_R\, +
\nonumber\\
&tr\left\{\Gamma_D^N \Gamma_D^S \mathcal P_{\frac{1}{2}}
\Gamma_D^i \Gamma_D^T\right\} s_S\,  t_T\,  (p-r)_N \, .
\label{first-s1}
\end{align}
We first consider the second and the third terms. We see that for $D$
odd these are zero, since $\mathcal P_{\frac{1}{2}}^\prime$ contains a
$\Gamma_{(D-1)}^{(D-1)+1}$ and the trace could be non zero only if all
the four $\Gamma$-matrices were of $\mu$-type.  For $D$ even fermions
can be chiral in which case there should be a corresponding insertion
of ($\mathbf{1}\pm\Gamma_D^{D+1}$) in the trace over
$\Gamma$-matrices. The terms containing $\Gamma_D^{D+1}$ vanish (using
the same argument as that of odd dimensions) and the remaining terms
are the ones present in Eq.~(\ref{first-s1}). The argument is reversed
if we use $\mathcal P_{\frac{1}{2}}^\prime$ but the final result will
remain unchanged so that without loss of generality we will consider
the projection $\mathcal P_{\frac{1}{2}}$. For $D\ge 10$ the trace is
zero, for the same reason it was zero in the one-loop case for $D\ge
8$. So we have to see what happens for $D=8$ (we do not need to
consider the case with $D=6$ since in this case there are
contributions already at one loop). In this case the trace turns out
to be proportional to $\epsilon^{ijkl}$ (modulo index permutations),
which is the completely antisymmetric tensor. Now we have to look at
the momenta: if we calculate the fluxes in all the considered cases we
find that these are symmetric in the indices $jkl$, so that also for
$D=8$ the contribution of the second and the third terms of
Eq.~(\ref{first-s1}) vanishes.  Now we move to the first term. This
can be rewritten in the following form:
\be
(2-D)\, tr\left\{\Gamma_D^S \mathcal P_{\frac{1}{2}} \Gamma_D^T\right\} 
s_S \, t_T \, (r+q)_i \, .
\label{first-s2}
\ee 
We immediately see that the trace we have to compute is the same as in
the one-loop case, so we can conclude that we can have a contribution
only for $D=6$.  Up to now we have computed diagrams in which the
twisted momentum is $s$. The computation of diagrams in which $t$ is
twisted is analogous and leads to the same result.\\

Now we consider the diagram on the right of Fig.~\ref{twoloops}. It is
proportional to: 
\begin{align}
&\sum_{\vec{q},\, \vec{s},\, \vec{q}^{\,\prime},\, \vec{s}^{\,\prime}}
\int\frac{d^4q}{(2\pi)^4}\frac{d^4s}{(2\pi)^4}
\frac{1}{q^2}\frac{1}{s^2}\frac{1}{t^2}\frac{1}{r^2}\,
tr\Big\{ \,/\!\!\!q \, \left(\delta_{\vec{q}-\vec{q}^{\,\prime}}+
\lambda_{\mathcal R}\otimes\mathcal P_{\frac{1}{2}} 
\delta_{\vec{q}+\vec{q}^{\,\prime}}\right)
\nonumber\\
&\Gamma_D^N \,T^B_{\mathcal R} 
\,\delta_{\vec{q}^{\,\prime}-\vec{s}-\vec{t}} \,/\!\!\!s 
\left(\delta_{\vec{s}-\vec{s}^{\,\prime}}+
\lambda_{\mathcal R}\otimes\mathcal P_{\frac{1}{2}}
\delta_{\vec{s}+\vec{s}^{\,\prime}}\right)\Gamma_D^R \,T^C_{\mathcal R}  
\,\delta_{\vec{s}^{\,\prime}+\vec{t}^{\,\prime}-\vec{r}^{\,\prime}} 
\, \delta^{BC}\, g_{NR}
\nonumber\\
&\left(\delta_{\vec{t}-\vec{t}^{\,\prime}}
+ \eta^C \alpha_R\, \delta_{\vec{t}+\vec{t}^{\,\prime}}\right) 
\,/\!\!\!r \,\left(\delta_{\vec{r}-\vec{r}^{\,\prime}}
+\lambda_{\mathcal R}\otimes\mathcal 
P_{\frac{1}{2}} \delta_{\vec{r}+\vec{r}^{\,\prime}}\right)
\Gamma_D^M \,T^A_{\mathcal R}
\delta_{\vec{p}-\vec{q}+\vec{r}}\Big\}\, ,
\label{seconddiagram}
\end{align}
where $t^\mu=q^\mu-s^\mu$ and $r^\mu=q^\mu-p^\mu=q^\mu$ and also here
we omit possible chirality projectors for chiral fermions in even
dimensions. Also in this case there are sixteen different diagrams
which can be grouped according to the number of twisted
fermions. Indeed the presence of a twisted gauge boson introduces a
sign ($\eta^C$), changes the flux of momenta and adds a factor
$(8-D)/D$~\footnote{In the case of no twist we have a factor $D$ due
to $\Gamma_D^N \Gamma_D^R g_{NR}$, while in presence of a twist this
changes to $\Gamma_D^N \Gamma_D^R g_{NR} \alpha_R = 8-D$.} with
respect to the case of no twist, but does not affect the structure of
the traces over Dirac and gauge indices.

We begin with diagrams with zero twisted fermions. These are
proportional to
\begin{equation}
\sum_{B,C}\delta_{BC}\, tr\left[T^A_{\mathcal
R}\left\{T^B_{\mathcal R},T^C_{\mathcal
R}\right\}\right]=\frac{1}{2}\sum_{B,C}\delta_{BC}\, \mathcal A\ d^{ABC}
\label{second-zerotf}
\end{equation}
where $\mathcal A$ is the four-dimensional anomaly coefficient. Since
spinors in the bulk of a $D$ dimensional space (whatever they are
chiral or Dirac with respect to the $\Gamma_D^{D+1}$ projection) are
made of four-dimensional Dirac spinors, the anomaly coefficient
$\mathcal A$, along with the tadpole, vanishes~\footnote{In fact a
Dirac fermion $(\eta,\bar\chi)^T$ in the complex representation
$\mathcal R$ is equivalent to Weyl spinors $\eta$ and $\chi$ in
representations $\mathcal R$ and $\bar{\mathcal R}$, respectively, and
the system is anomaly-free.}.

Diagrams with two twisted fermions can be related to
Eq.~(\ref{second-zerotf}) simply by using the commutation property
\begin{equation}
T^A_{\mathcal R}\lambda_{\mathcal R}=\eta^A \lambda_{\mathcal
R}T^A_{\mathcal R}
\label{comm-lambda}
\end{equation}
and therefore they also vanish.

We now consider diagrams with one twisted fermion. If $q$ is the
twisted momentum the amplitude is proportional to
\be
tr\left\{\barra q\,\mathcal P_{\frac{1}{2}} 
\barra s\,\barra r \,\Gamma_D^M\right\} \, .
\label{second-onetf}
\ee
Also in this case we compute it for $M=i$. We immediately observe that
we have reduced to the same situation of the second and the third
terms of Eq.~(\ref{first-s1}) and, in an analogous way, it can be
shown that also this diagram can be non zero only for $D=6$. If the
twisted momentum is $s$ or $r$, these diagrams differ from the one
discussed above simply by a permutation and therefore we obtain the
same result.

The diagram with three twisted fermions can be related to
Eq.~(\ref{second-onetf}) by taking into account the commutation
property of $\mathcal P_{\frac{1}{2}}$
\begin{equation}
\Gamma_D^M \mathcal P_{\frac{1}{2}}=
\pm \, \alpha_M \mathcal P_{\frac{1}{2}}\Gamma_D^M \, ,
\label{comm-P}
\end{equation}
where the ``$-$'' sign corresponds to $\mathcal
P_{\frac{1}{2}}^\prime$, as well as the property $\mathcal
P_{\frac{1}{2}}^2=1$. Therefore we conclude that also this one can be
non-zero only for $D=6$.

\section{\sc Conclusions and outlook}
\label{Conclusions}
In orbifold theories with Higgs-gauge unification Higgs fields $A_i$
are internal components of higher-dimensional gauge fields and as such
they transform in the vector representation of the tangent space
rotation group $SO(d)$. A quadratically divergent mass term for Higgs
fields in the bulk is forbidden by the higher-dimensional gauge
invariance while the remnant shift symmetry allows for a similar term
localized at the orbifold fixed points only through the non-abelian
component of a $U(1)$ field strength tadpole. We have obtained the
general conditions an orbifold gauge theory must fulfill for the
absence of tadpoles localized at fixed points where $U(1)$ factors are
left over by the orbifold projection.  On the one hand the internal
rotation group at an orbifold $\mathbb G_f$-fixed point is defined as
the $SO(d)$ subgroup commuting with $\mathbb G_f$.  On the other hand
the localized tadpoles can only appear as the invariant terms
$\epsilon^{ij}F_{ij}$, where $\epsilon^{ij}$ is the Levi-Civita
connection which transforms covariantly under the rotation group
$SO(2)\subseteq SO(d)$.  In this way the existence of tadpoles is tied
to the possibility that $SO(2)$ be a subgroup factor of the internal
rotation group.  Stated differently, the rotation group acting on some
internal indices being $SO(p)$ ($p>2$) is enough to guarantee the
absence of tadpoles involving the corresponding components. A
particularly simple example is provided by $T^d/\mathbb Z_2$ orbifolds
where the internal rotation group for all fixed points is
$SO(d)$. There the absence of tadpoles is guaranteed for theories with
$d>2$.

In this paper we did not attempt to construct a realistic Higgs-gauge
unification model but only to fix the general conditions for the
absence of tadpoles and providing general examples where these
conditions are fulfilled. A number of problems should be solved before
a realistic model can be drawn. First of all we must consider models
with $D>6$ dimensions. In fact $D=5$ models were first studied and
they generically lead to too low Higgs masses due to the absence of
quartic terms in the potential~\footnote{One can avoid this
shortcoming by introducing an extra scalar but in this case one
reintroduces the problem of quadratic divergences or else one must
reintroduce supersymmetry~\cite{Burdman:2002se}.}. $D=6$ models have a
quartic coupling from gauge interactions but it was proven that they
generically lead to UV sensitivity through the localized
tadpoles~\footnote{A possible way out is given by the possibility that
the local tadpoles are globally vanishing although this leads to
strong constrains in the bulk fermion spectrum~\cite{Scrucca:2003ut}.}
except for $\mathbb Z_2$ models with only bulk gauge fields. Since
bulk fermions are generically required to trigger electroweak symmetry
breaking we are thus led to consider models with $D>6$ dimensions. The
trivial examples we presented in this paper predict the existence of
$d$ Higgs fields leading to non-minimal models. Of course the
conditions that preclude the existence of quadratic divergences for
Higgs fields do not forbid the radiative generation of finite $\sim
(1/R)^2$ masses, that vanish in the $R\to\infty$ limit. Some of the
above Higgs fields can acquire different masses and even not
participate in the electroweak symmetry breaking phenomenon, depending
on the models.

Another problem that we are not facing here is the generation of
fermion masses. If fermions are localized at orbifold fixed points
they can develop Yukawa couplings through Wilson line interactions
after the heavy bulk fermions have been integrated
out~\cite{Csaki:2002ur,Scrucca:2003ra}. One could think of localizing
bulk fermions similarly to the $D=5$ case~\cite{localization} by
giving them a bulk mass~\cite{Lee:2003mc}. Otherwise the bulk fermions
acquire a mass dictated by the higher-dimensional bulk gauge coupling
$\sim g_D/(\pi R)^{d/2} v$: if this coupling is large (say comparable
to the top quark Yukawa coupling) it can trigger an efficient
electroweak symmetry breaking. In this case the four-dimensional gauge
couplings should be dominated by localized gauge kinetic terms
corresponding to the unbroken subgroups left over at the orbifold
fixed points.

\vspace*{7mm}
\subsection*{\sc Acknowledgments}

\noindent We acknowledge discussions with C.~Scrucca, M.~Serone and
A.~Wulzer. We would like to thank the CERN theoretical division, where
part of this work was done, for kind hospitality. This work was
supported in part by the RTN European Programs HPRN-CT-2000-00148 and
HPRN-CT-2000-00152, and by CICYT, Spain, under contracts FPA 2001-1806
and FPA 2002-00748.

\vspace*{7mm}

\appendix

\section{\sc $\Gamma$-matrices and traces in arbitrary dimension}
\label{Gamma}

We work with the metric mostly $-1$. For the 4D $\gamma$-matrices we
adopt the following convention:
\be
\label{gammamu_4D}
\gamma^{\mu} =
\left(\begin{array}{cc}
0 & \sigma^{\mu} \\
\bar{\sigma}^{\mu} & 0
\end{array}\right)\ ,
\ee
with $\sigma^{\mu}=(\mathbf{1},\vec{\sigma})$ and
$\bar{\sigma}^{\mu}=(\mathbf{1},-\vec{\sigma})$ and we define
$\gamma^5$ as:
\be
\label{gamma5_4D}
\gamma^5 = i\gamma^0 \gamma^1 \gamma^2 \gamma^3 =
\left(\begin{array}{cc}
-\mathbf{1} & 0\\
0 & \mathbf{1}
\end{array}\right)\ .
\ee
In $D=2n$ dimensions the $\Gamma$-matrices $\Gamma_D^M$ are defined
recursively in this way:
\be
\label{gammaM_DD}
\begin{array}{l}
\Gamma^M_D = \mathbf{1}_{2\times2}\otimes \Gamma^M_{D-2}~~~
M=\mu,5,\dots,D-2\\[0.2cm]
\Gamma^{D-1}_D = i \sigma_1 \otimes \Gamma^{(D-2)+1}_{D-2}\\[0.2cm]
\Gamma^D_D = -i \sigma_2 \otimes \Gamma^{(D-2)+1}_{D-2}\ .
\end{array}
\ee
$\Gamma^{D+1}_D$ is defined by:
\be
\label{gamma5_DD}
\Gamma^{D+1}_D = (-1)^{\frac{D-2}{4}} \Gamma^0_D \Gamma^1_D \cdots\Gamma^D_D = 
- \sigma_3 \otimes \Gamma^{(D-2)+1}_{D-2}   \ .
\ee
In $D=2n+1$ dimensions the first $(D-1)$ $\Gamma^M_D$ coincide with
$\Gamma^M_{D-1}$, while $\Gamma^D_D = i \Gamma^{(D-1)+1}_{D-1}$.  It
is straightforward to verify that these $\Gamma$-matrices satisfy the
correct anticommutation rules.\\

We can now proceed to the evaluation of the traces of these
$\Gamma$-matrices. We list here the results in the case of $D$
even. The case with odd dimensions can be recovered simply remembering
that the $\Gamma_{2n+1}$-matrices coincide with the $\Gamma_{2n}$,
except for $\Gamma_{2n+1}^{2n+1}=i\,\Gamma_{2n}^{2n+1}$. Then the
traces in the odd case can be obtained from the even one, paying
attention to the presence or not of $\Gamma_{2n+1}^{2n+1}$. For $D$
even we have:
\be
\label{oddgamma}
tr\{\textrm{odd } \# \textrm{ of } \Gamma^M_D\}=0
\ee
\be
\label{2gamma}
tr\{\Gamma^M_D\Gamma^N_D\}= 2^{\frac{D}{2}}g^{MN}
\ee
\be
\label{4gamma}
tr\{\Gamma^M_D\Gamma^N_D\Gamma^P_D\Gamma^Q_D\}= 
2^{\frac{D}{2}}(g^{MN}g^{PQ}-g^{MP}g^{NQ}+g^{MQ}g^{NP})
\ee
\be
\label{6gamma}
tr\{\Gamma^M_D\Gamma^N_D\Gamma^P_D\Gamma^Q_D\Gamma^R_D\Gamma^S_D\}= 
2^{\frac{D}{2}}(g^{MN}g^{PQ}g^{RS}-\cdots)
\ee
and so on.  If $\Gamma^{D+1}_D$ is involved, the trace is always
zero unless $\Gamma^{D+1}_D$ is multiplied by $D$ $\Gamma^M_D$ with
all the $M$ different; it is precisely:
\be
\label{Dgamma+gamma5}
tr\{\Gamma^{M_1}_D...\Gamma^{M_D}_D\Gamma^{D+1}_D\}=
- (-1)^{\frac{D-2}{4}}2^{\frac{D}{2}}\epsilon^{M_1...M_D}=
i(2i)^{\frac{D}{2}}\epsilon^{M_1...M_D}\ .
\ee
%


\end{document}